\documentclass[aip, jcp, preprint]{revtex4-2}
\usepackage{graphicx} 
\usepackage{color}
\usepackage{xspace}
\usepackage{multirow}
\usepackage{mhchem}
\usepackage{subfig}
\usepackage{booktabs}
\usepackage{soul}
\usepackage{chemformula}
\usepackage{comment}

\newcommand{\pyscf}{\textsc{PySCF}\xspace}
\newcommand{\gpupyscf}{\textsc{GPU4PySCF}\xspace}
\newcommand{\mpipyscf}{\textsc{MPI4PySCF}\xspace}
\newcommand{\pyscfad}{\textsc{PySCFAD}\xspace}
\newcommand{\numpy}{{NumPy}\xspace}
\newcommand{\scipy}{{SciPy}\xspace}
\newcommand{\jax}{\textsc{JAX}\xspace}

\begin{document}
\title{The Python Simulations of Chemistry Framework: 10 years of an open-source quantum chemistry project}

\author{Qiming Sun}
\affiliation{Quantum Engine LLC, Lacey, WA 98516, USA}

\author{Matthew R Hermes}
\affiliation{Department of Chemistry and Chicago Center for Theoretical Chemistry, University of Chicago, Chicago, IL 60637, USA}

\author{Xiaojie Wu}
\affiliation{Independent Researcher, Santa Clara, CA 95051, USA}

\author{Huanchen Zhai}
\affiliation{Initiative for Computational Catalysis, Flatiron Institute, New York, NY 10010, USA}
\affiliation{Division of Chemistry and Chemical Engineering,
	California Institute of Technology, Pasadena, CA 91125, USA}

\author{Xing Zhang}
\affiliation{Division of Chemistry and Chemical Engineering,
	California Institute of Technology, Pasadena, CA 91125, USA}
\affiliation{Marcus Center for Theoretical Chemistry,
	California Institute of Technology, Pasadena, CA 91125, USA}

\author{Abdelrahman M. Ahmed}
\affiliation{Department of Chemistry and Biochemistry, The Ohio State University, Columbus, OH 43210, USA}

\author{Juan José Aucar}
\affiliation{Department of Physics, National University of Northeast Argentina, Corrientes, W3404AAS, Argentina}

\author{Oliver J. Backhouse}
\affiliation{Department of Physics, King's College London, London WC2R 2LS, UK}

\author{Samragni Banerjee}
\affiliation{Department of Chemistry and Biochemistry, The Ohio State University, Columbus, OH 43210, USA}

\author{Peng Bao}
\affiliation{Beijing National Laboratory for Molecular Sciences, State Key Laboratory for Structural Chemistry of Unstable and Stable Species, Institute of Chemistry, Chinese Academy of Sciences, Zhongguancun, Beijing 100190, China}


% ...

\begin{abstract}
Over the past decade, the Python-based Simulations of Chemistry Framework (PySCF) has developed into a widely used open-source platform for electronic-structure theory and quantum chemical method development. This article reviews the major advances since the previous overview in 2020,
covering new modules and methodology, infrastructure changes, and performance benchmarks.
\end{abstract}

\maketitle

\section{Introduction}

The Python based Simulations of Chemistry Framework (\pyscf)\cite{sun2018pyscf,sun2020recent} is a widely used quantum chemistry library. Initially founded as an internal project within the Chan group, the first stable public release (version 1.0) was in 2015, and the version at the time of writing (version 2.12) marks ten years of this open-source project. Currently, the \pyscf repositories encompass more than {500,000} lines of code, with more than 1,000 dependent projects on GitHub, and the package receives more than 1,000,000 downloads a year, making it among the most widely used development frameworks in quantum chemistry.

The original intention behind \pyscf was to create a framework that supported not only the traditional quantum chemistry paradigm where the user supplies an input file and reads the calculation output, but also the development of new quantum chemistry capabilities. To this end, inspired by the ubiquitous \numpy\cite{numpy} and \scipy\cite{scipy} libraries used in numerical computation in Python, \pyscf is structured as a set of loosely coupled modules with small, largely pure, functions that implement a rich set of APIs. The implementation is mainly in Python with performance hotspots rewritten in C. In addition, in part due to the initial applications that motivated the development of \pyscf (as recounted in Ref.~\citenum{sun2020recent}), the code is designed to support molecular, materials, and model Hamiltonians on a (relatively) equal footing, at both the mean-field and many-body level. The flexibility of this design has contributed to \pyscf being adopted in areas outside of the chemistry community: materials science, condensed matter physics, machine learning, and quantum information science. The usage of \pyscf has thus moved far beyond the interests and expertise of the original developers. Fortunately, to a large degree, the original design choices have held through this expansion. 

As we look to the next decade of this project, we see user demands for new kinds of calculations (e.g. high throughput for dataset generation); simulations of ever  larger systems; and the emergence of new computing hardware, and programming libraries and paradigms, and with it, a growing complexity of management. 
In this work, we document the main developments in \pyscf since the last publication (at the time of version 1.7)~\cite{sun2020recent} that aim to address some of these emerging challenges, and provide some outlook on the future.

\section{Main developments}

\subsection{Code organization}

Since Ref.~\citenum{sun2020recent}, the codebase has greatly expanded. In addition to  new methods (and new implementations of existing methods), significant effort has been made to optimize for hardware (especially GPUs) and to encompass new development paradigms, for example,  auto-differentiation and just-in-time (JIT) compilation frameworks. 

Because of this expansion, \pyscf is now separated into the core \texttt{pyscf} repository, which provides a compact set of modules that can be expected to be used in the widest range of applications; the \texttt{pyscf-forge} repository, which provides a staging ground for experimental features that may later be transitioned to the core \texttt{pyscf} repository; as well as companion repositories such as \gpupyscf, \pyscfad, and \mpipyscf. Finally, \pyscf has a native mechanism to support custom extensions, and a list of extensions developed by the community is maintained on the \pyscf website, including links to their independent repositories. Unlike those in \texttt{pyscf} and \texttt{pyscf-forge}, the features available as extensions are not officially supported by the developers.



\subsection{Periodic electronic structure infrastructure}

A key design principle of \pyscf is that it should support both molecular and materials [periodic boundary condition (PBC)] calculations on a relatively equal footing. 
To this end, \pyscf implements multiple mathematical formalisms to support periodic electronic structure calculations (mean-field and beyond) in Gaussian basis sets. The different formalisms are primarily based on different decompositions of the periodic electron-electron repulsion integrals via the density-fitting (DF) approximation, using either an auxiliary basis of plane waves (FFTDF) or atom-centered Gaussians (GDF). As these have different performance characteristics, users can select the optimal choice depending on their application.

\begin{figure}[]
\begin{center}
	\includegraphics{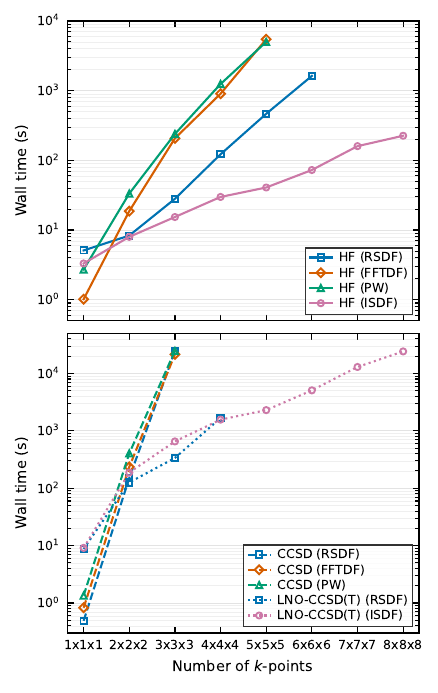}
\end{center}
\caption{
    Wall times of ground-state HF (upper), and CCSD and LNO-CCSD(T) correlation energy (lower) calculations for the diamond primitive cell using the GTH-cc-pVDZ\cite{ye2022correlation} Gaussian basis set with various density fitting schemes: RSDF (blue), FFTDF (orange), and ISDF (purple), as well as a plane wave basis (green). A plane wave energy cutoff of 60 a.u. and the GTH-HF-rev\cite{stein2020double} pseudopotential were used throughout. The calculations were performed on 32 Intel Ice Lake 8352Y CPU cores [for LNO-CCSD(T) and ISDF] and on 32 AMD EPYC 7742 CPU cores (for all other methods).
}
\label{fig:df_time}
\end{figure}

Since Ref.~\citenum{sun2020recent}, multiple enhancements have been made to the FFTDF and GDF infrastructure in \pyscf to improve performance and accuracy. 
The improvements to the FFTDF module now make it suitable for use in a wide range of applications so long as pseudopotentials are acceptable. For example, the multigrid FFTDF density functional theory (DFT) implementation now achieves state-of-the-art performance on both CPUs and GPUs (see Table~\ref{tab:multigrid}), and
the GPU parallelization can saturate the compute of modern NVIDIA A100 and H100 GPUs. Using multigrid FFTDF, DFT iterations on systems with as many as 20000 basis functions now take only a few seconds on a single GPU. Similarly, using the occ-RI {and interpolative separable density fitting (ISDF)} techniques,\cite{sharma2022fast,yang2026abinitiobodyquantum} Hartree-Fock (HF) and hybrid DFT calculations can also now be carried out with good performance using FFTDF on both CPUs and GPUs (see Fig.~\ref{fig:df_time}).

\begin{table*}
	\caption{
		Wall times (in seconds) for one self-consistent field (SCF) iteration
		and nuclear gradient in multigrid FFTDF restricted KS-DFT calculations for water clusters, using the
		PBE\cite{pbe,libxc} density functional, GTH-TZV2P\cite{vandevondele2007gaussian} basis set and GTH-PADE\cite{hartwigsen1998relativistic} pseudopotential. 
        A plane wave cutoff energy of 140 a.u. was used.
        The calculations were performed on NVIDIA A100 (80 GB) and H100 (80 GB) GPUs, as well as on Intel Cascade Lake 8276 CPUs.
	}\label{tab:multigrid}
	\scriptsize
	\centering
	\begin{tabular*}{1.0\textwidth}{@{\extracolsep{\fill}}rrcccccc}
		\hline\hline
		&&\multicolumn{3}{c}{1 SCF iteration on average} & \multicolumn{3}{c}{Nuclear gradient}  \\
		\cline{3-5}\cline{6-8}
		\multirow{2}{*}{System} & \multirow{2}{*}{$N_\text{basis}$}  
        & \multicolumn{2}{c}{\gpupyscf} & \pyscf 
        &\multicolumn{2}{c}{\gpupyscf} & \pyscf \\
        \cline{3-4} \cline{5-5} \cline{6-7} \cline{8-8}
		& & H100 & A100 & CPU (28 cores) & H100 & A100 & CPU (28 cores) \\
        \hline
		32  \ce{H2O} &   1280 &   0.06 &   0.08 &   0.52 &   0.25 &   0.39 &   1.25 \\
		64  \ce{H2O} &   2560 &   0.12 &   0.20 &   1.83 &   0.42 &   0.69 &   3.33 \\
		128 \ce{H2O} &   5120 &   0.41 &   0.77 &   8.97 &   1.08 &   1.96 &   9.03 \\
		256 \ce{H2O} &  10240 &   1.86 &   3.67 &  52.52 &   3.85 &   7.47 &  29.37 \\
		512 \ce{H2O} &  20480 &  13.48 &  31.20 & 226.28 &  15.58 &  37.46 & 108.30 \\ 
        \hline\hline
    \end{tabular*}
\end{table*}

GDF is intended to support all-electron basis sets and is also a good option for hybrid and exact exchange mean-field calculations. To enhance the efficiency of such
calculations, range-separated integral evaluation techniques
have been developed in \pyscf. The range-separated Gaussian density fitting (RSGDF)\cite{ye2021fast}
method provides a significant acceleration compared to the original
GDF implementation {for large unit cells} enabling the efficient treatment
of periodic hybrid DFT and post-HF methods at the $\Gamma$ point and for
systems with a moderate number of $k$-points (typically up to $\sim 100$, see Fig.~\ref{fig:df_time}). To further
extend scalability, a range-separated Coulomb and exchange matrix construction
algorithm {(RSJK)} \cite{sun2023exact,sun2023efficient} has been introduced, which is an integral-direct implementation without DF, allowing hybrid DFT calculations {on modest sized unit cells} but with more than 10,000 $k$-points to be performed efficiently.\cite{sun2023exact}

Finally, in addition to Gaussian basis set periodic electronic structure, \pyscf now has a fully featured pure plane-wave basis infrastructure for three-dimensional periodic systems (currently in \texttt{pyscf-forge}). The supported mean-field methods are HF and DFT (with LDA, GGA, meta-GGA, and global hybrid functionals), in both spin-restricted and spin-unrestricted forms. \pyscf's existing Davidson algorithm is used as the eigensolver, and effective potential mixing is used to improve convergence. All mean-field calculations support space group symmetry and occupation smearing. For post-HF plane-wave calculations, restricted and unrestricted MP2 are available, along with restricted CCSD. To aid in basis set convergence, utilities are provided to convert virtual orbitals from a plane-wave basis to a Gaussian basis before performing post-HF calculations.~\cite{boothPlaneWavesLocal2016} GTH~\cite{goedeckerSeparableDualspaceGaussian1996} and ccECP~\cite{bennettNewGenerationEffective2017} pseudopotentials are supported for all calculations, and SG15 pseudopotentials~\cite{hamannOptimizedNormconservingVanderbilt2013,schlipfOptimizationAlgorithmGeneration2015} are experimentally supported for mean-field calculations.
A comparison of the performance of plane-wave and Gaussian basis set Hartree-Fock implementations is shown in Fig.~\ref{fig:df_time}. 

\subsection{Density functional theory and related methods}

\pyscf has comprehensive support for DFT calculations in both molecules and materials. 
In recent years, DFT functionality has been updated with new exchange-correlation (XC) treatments
and additional properties. The analytical derivatives of the VV10\cite{vydrov2010nonlocal} nonlocal correlation functional have been derived up to second order, which enables the
computation of analytic nuclear gradients and Hessians, as well as
polarizabilities and related properties.
The DFT modules have also been refactored to better support calculations with DFT-D3\cite{grimme2010consistent} and DFT-D4\cite{caldeweyher2017extension} dispersion corrections.

The challenges of non-collinear functionals are addressed by the
multi-collinear DFT framework\cite{pu2023noncollinear} in \pyscf. The multi-collinear XC module fully
supports energy evaluation and higher-order derivatives. Non-collinear computations can be performed within general Kohn-Sham (GKS) theory with spin-orbit coupling (SOC), as well as in two-component and four-component relativistic Dirac-Kohn-Sham calculations. Higher-order derivatives for
multi-collinear XC functionals have been integrated into the time-dependent density functional theory (TDDFT) code. Other TDDFT features, including the spin-flip TDDFT variant and
derivative coupling computations, have also been implemented in \pyscf.

\begin{table*}
	\caption{
		Wall times (in seconds) for one SCF iteration on average
		and nuclear gradient computation in restricted all-electron HF calculations of linear alkane chains using various exchange algorithms. ``Analytical'' indicates the exact evaluation of the four-center integrals for the exchange potential (K-matrix), ``DF'' indicates density fitting with the def2-universal-jkfit basis,~\cite{Weigend2008hartree} and ``SGX'' refers to seminumerical exchange. For consistency, the Hartree potential (J-matrix) was computed using density fitting with the def2-universal-jkfit basis for all calculations, so the only difference between methods is the choice of exchange algorithm. Each calculation was performed on a 96-core AMD Genoa node.
	}\label{tab:sgx}
	\scriptsize
	\centering
    \begin{tabular*}{1.0\textwidth}{@{\extracolsep{\fill}}crcccccc}
        \hline\hline
        Number of & \multirow{2}{*}{Basis} &\multicolumn{3}{c}{1 SCF iteration on average} & \multicolumn{3}{c}{Nuclear gradient}  \\
        \cline{3-5}\cline{6-8}
        Carbons &&
        Analytical & \textsc{DF} & \textsc{SGX} &
        Analytical & \textsc{DF} & \textsc{SGX}\\\hline
        20 &   def2-TZVP & 2.8 & 1.5 & 3.2 & 25.8 & 9.3 & 13.6 \\
        40 &   def2-TZVP & 9.1 & 9.1 & 8.6 & 96.7 & 72.1 & 46.1 \\
        80 &   def2-TZVP & 33.2 & 89.2 & 26.2 & 422.0 & 669.7 & 193.6 \\\hline
        20 & def2-QZVPPD & 71.8 & 10.4 & 15.5 & 941.3 & 45.0 & 79.2 \\
        40 & def2-QZVPPD & 323.3 & 96.0 & 58.2 & 4137.3 & 372.1 & 352.0 \\\hline\hline
    \end{tabular*}
\end{table*}

Seminumerical (or pseudospectral) exact exchange algorithms~\cite{friesnerSolutionSelfconsistentField1985,neeseEfficientApproximateParallel2009,laquaEfficientLinearScalingSeminumerical2018,helmich-parisImprovedChainSpheres2021} accelerate the computation of the exact exchange operator by evaluating the integration over one coordinate of the two-electron integrals analytically and the other numerically. The seminumerical exchange program in \pyscf has recently been accelerated, with efficient grids,~\cite{helmich-parisImprovedChainSpheres2021} modified Becke partitioning,~\cite{laquaImprovedMolecularPartitioning2018} and an improved density matrix screening algorithm to achieve reduced asymptotic scaling for large systems. The speed and precision of the density matrix screening can be balanced by tuning intuitive parameters for the energy and potential error tolerance. Example timings compared to analytical and DF exchange are shown in Table~\ref{tab:sgx}.

The scope of molecular and materials property computations has also been broadened
at the DFT level. Newly developed features enable the evaluation of infrared (IR) and Raman spectra, energy decomposition analysis (EDA),\cite{sharma2018active} and four-component parity-violating contributions.\cite{berger2019parity,aucar2023relationship}
For periodic systems, \pyscf  supports the
computation of dipole moments, polarizabilities, and hyperpolarizabilities.
Relativistic corrections for all-electron DFT calculations under PBCs are also now supported. These corrections are treated with the exact two-component (X2C) framework,\cite{yeh2022relativistic} with both spin-free and spin-orbit coupling formulations available.
Nuclear gradient and stress tensor evaluations are  available for both DFT and DFT+U (see below) methods for solids. These
functionalities are easily accessible with a provided \texttt{Calculator} for the Atomic Simulation Environment~\cite{larsen2017atomic} to enable geometry and crystal structure optimizations.


Finally, \pyscf  includes a DFT+U framework for both molecules and solids.
In periodic calculations, both $\Gamma$-point and $k$-point sampling are supported. The Hubbard U parameters can be obtained using the linear-response
DFT+U method\cite{Cococcioni2005} in both molecular and periodic modules, enabling a systematic way to compute system-specific U values. 

\subsection{High-order and low-scaling correlated wavefunction quantum chemistry}

\begin{figure*}
\includegraphics[width=\textwidth]{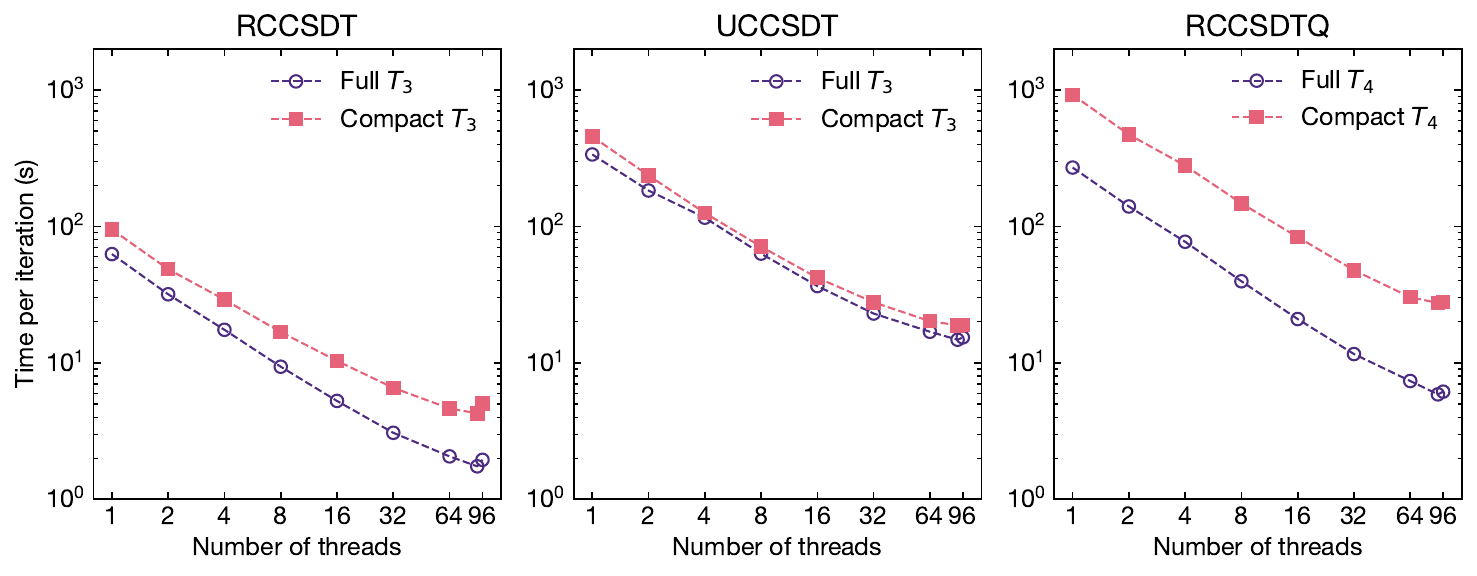}
\centering
\caption{Average per-iteration wall times as a function of the number of threads for RCCSDT, UCCSDT, and RCCSDTQ calculations. All calculations were performed for the hydrogen thioperoxide (\ce{H2OS}) molecule within the frozen-core approximation, using the cc-pVTZ basis for RCCSDT and UCCSDT ($N_{\text{occ}} = 7$, $N_{\text{vir}} = 79$) and the cc-pVDZ basis for RCCSDTQ ($N_{\text{occ}} = 7$, $N_{\text{vir}} = 29$). ``Full'' refers to implementations that explicitly store the complete $T_3$ (RCCSDT and UCCSDT) or $T_4$ (RCCSDTQ) amplitude tensors, while ``Compact'' refers to implementations employing compact tensor storage and contraction by utilizing index-permutation symmetry. All benchmarks were run on a single 96-core AMD Genoa node. Tensor contractions were performed using \texttt{pytblis},\cite{pytblisrepo} a Python wrapper for the \texttt{TBLIS} library.\cite{tblisrepo}}
\label{fig:highordercc}
\end{figure*}

Since its beginning, \pyscf has had robust support for standard correlated methods of quantum chemistry, namely M{\o}ller-Plesset perturbation theory (such as MP2), configuration interaction theory, and coupled-cluster (CC) theory. 
In the latest version, \pyscf includes efficient implementations of CCSDT (spin-restricted and unrestricted) and CCSDTQ (restricted) energy calculations. These implementations use a $t_1$-dressed formalism~\cite{Koch1994} along with a lower-triangular storage of amplitudes. The spin summation~\cite{Springer2019} and tensor index permutation are implemented in C code with shared-memory parallelization, while tensor contractions are written in pure Python for readability and easy modification. External tensor contraction libraries, such as TBLIS,\cite{Matthews2018,Huang2018} improve the performance when available. The spin-restricted implementations use the non-orthogonal spin-adaptation approach.\cite{Matthews2013,Matthews2015} Limited primarily by memory costs, CCSDT and CCSDTQ calculations can be performed for systems with about {400} orbitals and {80} orbitals, respectively (assuming {30} electrons and 1.5~TB memory). Representative timings as a function of the number of threads are shown in Fig.~\ref{fig:highordercc}.

In addition, there has been substantial development of lower scaling approximations to correlated wavefunction theories. 
The textbook implementation of these methods exhibits an unphysically high polynomial scaling due to their use of delocalized, canonical molecular orbitals. 
\pyscf now includes a cluster-in-molecule based approximation to CC theory, wherein many independent but overlapping fragment problems are solved and the total correlation energy is reconstructed. Specifically, \pyscf implements the local natural orbital (LNO) CCSD and CCSD(T) methods~\cite{rolik2011general,rolik2013efficient}, which are currently available in \texttt{pyscf-forge}. In this approach, each localized orbital defines a separate fragment, which is supplemented by a truncated set of occupied and unoccupied LNOs that diagonalize the respective orbital-specific density matrices (orbitals can also be combined into fragments). Within each fragment space, the CC equations are solved using the existing, canonical \pyscf implementations. The LNO eigenvalue threshold controls the accuracy and cost. Currently, LNO-CCSD(T) calculations can be performed for systems containing about 5,000 orbitals,~\cite{ye2023ab,ye2024adsorption,ye2024periodic} {and this limit can be pushed further by employing the ISDF technique\cite{yang2026abinitiobodyquantum} (see Fig.~\ref{fig:df_time})}. Although not yet in \pyscf, a recent work reported a low-scaling \pyscf implementation of MP2 and the random-phase approximation using the domain based localized pair natural orbital (DLPNO) approximation~\cite{liang2025efficient}; for combinations and comparisons of DLPNO and LNO methods, we refer to several previous works.
\cite{ye2023ab,ye2024adsorption,ye2024periodic,lnoccad,song2025random,yang2026abinitiobodyquantum}

\begin{figure}
\includegraphics{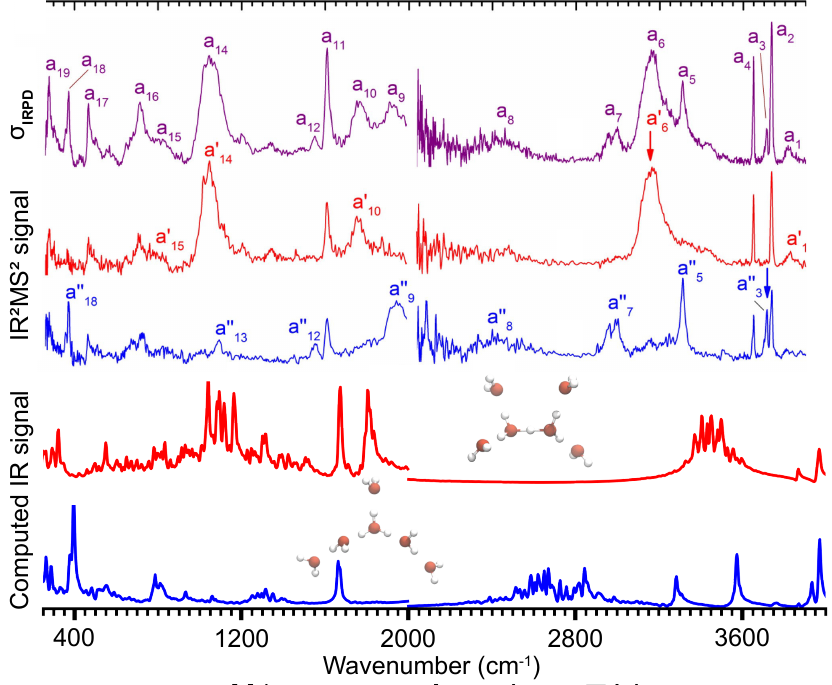}
\centering
\caption{IR spectra of the protonated water cluster. The first panel shows the experimental gas-phase \ch{H2}-predissociation spectrum of \ch{H^+(H2O)6.H2}. The second and third panels show the experimental IR$^2$MS$^2$ spectra of \ch{H^+(H2O)6.H2}, probing the transitions at 3159 cm$^{-1}$ and 3715 cm$^{-1}$ respectively (indicated by the red and blue arrows). 
(The experimental spectra were reprinted with permission from Ref.~\citenum{heine2013isomer}. Copyright {2013} American Chemical Society).
The last two panels show the computed IR spectra for the Zundel-like and Eigen-like conformers at the LNO-CCSD(T)/cc-pVTZ level of theory. The intensity under 2000 cm$^{-1}$ in the computed spectra is multiplied by 3 for clarity, and the spectra are convoluted using a Gaussian kernel with a width of 1 cm$^{-1}$. Reprinted with permission from Ref.~\citenum{lnoccad}. Copyright 2024 AIP Publishing.}
\label{fig:lnocc_ir}
\end{figure}

These local methods are available for molecules and periodic solids with or without $k$-point sampling.\cite{ye2024periodic} Moreover, as facilitated by the \pyscf design philosophy, the LNO infrastructure has already been repurposed to be automatically differentiable, allowing access to CCSD(T) gradients and response properties for large systems (see Fig.~\ref{fig:lnocc_ir}),\cite{lnoccad} and to develop a low-scaling implementation of auxiliary-field quantum Monte Carlo,\cite{kurian2023toward} the general framework of which is described next.

\subsection{Auxiliary field quantum Monte Carlo}

The latest version of \pyscf introduces support for auxiliary-field quantum Monte Carlo (AFQMC), a flavor of projector Monte Carlo methods that can provide accurate estimates of the electronic ground-state energy of a second-quantized Hamiltonian. 
The implementation of AFQMC within the same computational framework as other correlated wavefunction theories, such as
CI, MP, and CC theory, enables a direct comparison between the methods as well as a mixing of the methods, such as the use of correlated wavefunction trial states in AFQMC.

AFQMC typically begins with a HF trial state and employs imaginary-time propagation to project out contributions from excited eigenstates of the Hamiltonian. In the long imaginary-time limit, this procedure yields the ground state. The propagation is performed stochastically. Carrying out this projection in an unbiased manner is, however, infeasible for realistic systems due to the fermionic sign problem. This issue is addressed through the phaseless approximation,\cite{zhang2003quantum} which suppresses the sign problem at the cost of introducing a systematic bias in the ground-state energy. The magnitude of this bias depends on the quality of the chosen trial wavefunction. In the limit where the trial wavefunction approaches the exact ground state, the bias vanishes and AFQMC recovers the exact energy. For details of the algorithm, we refer the reader to Ref.~\citenum{mahajan2022selected}.

The HF state is the most commonly used trial wavefunction in AFQMC. In this case, the computational cost of the method scales as 
$O(N^5)$ with system size $N$, for fixed stochastic error. The resulting accuracy is typically between that of CCSD and CCSD(T).\cite{lee2022twenty} More recently, some of the \pyscf developers have introduced efficient algorithms for incorporating improved trial states,\cite{mahajan2025beyond} including those obtained from selected CI and systematically improvable CI expansions such as CISD, CISDT, etc. When a CISD wavefunction is used as the trial state, the computational scaling increases to 
$O(N^6)$ for fixed stochastic error. However, the resulting AFQMC predictions are significantly more accurate, often exceeding the quality of CCSD(T), even for organic molecules at equilibrium geometries, where CCSD(T) is widely regarded as the ``gold standard'' of quantum chemistry.

\begin{figure}
\includegraphics[width=0.9\textwidth]{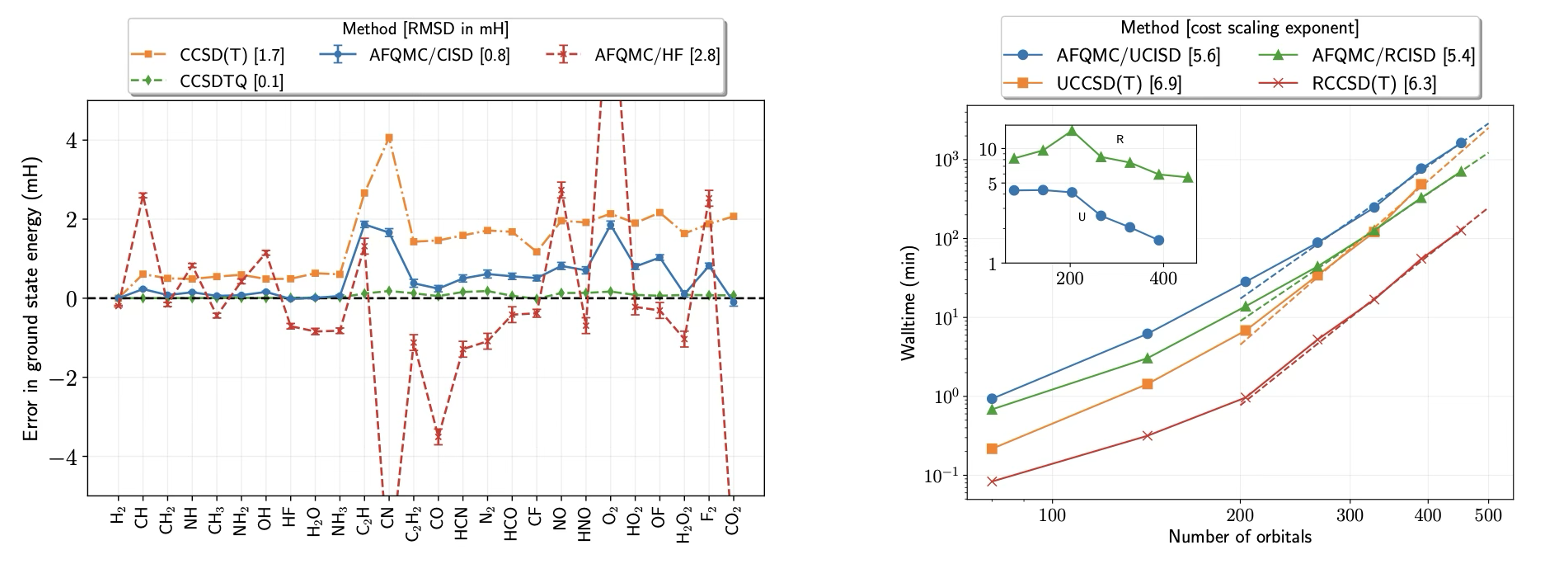}
\centering
\caption{(a) The left graph shows the error in energies of molecules in the HEAT dataset~\cite{tajti2004heat} for various methods relative to CCSDTQP results. AFQMC with a CISD trial state is more accurate than CCSD(T) for all molecules and it shows significant improvement over the results when a HF trial state is used. (b) The right graph shows the walltime of performing calculations on trans-polyacetylene of increasing size with AFQMC/CISD (GPU-accelerated) and CCSD(T) (CPU-based) using the aug-cc-pVDZ basis. Both restricted and unrestricted formalisms are shown for increasing system lengths. AFQMC sampling was scaled linearly with size to maintain a constant stochastic error of $\sim$1 m$E_\text{h}$. The inset displays the relative walltime ratios between the two methods. (Copyright 2025 ACS Publishing.)}
\label{fig:afqmc_t_a}
\end{figure}

\pyscf currently supports several variants of AFQMC, summarized below:
\begin{enumerate}
    \item \textbf{Mean-field trial wavefunctions:} AFQMC can be performed using restricted, unrestricted, or generalized mean-field trial states. The accuracy of the results typically improves as more symmetries are broken, at the cost of higher computational cost (the scaling of the method remains unchanged but the prefactor increases). AFQMC walkers themselves are symmetry-adapted and are often eigenfunctions of the $S^2$ operator, with $\langle S_z\rangle=(N_\alpha - N_\beta)/2$. This allows one to target specific spin sectors by initializing walkers with the appropriate numbers of $\alpha$ and $\beta$ electrons. AFQMC calculations can be performed for systems with up to $1000$ orbitals and the main bottleneck is the memory required to store two-electron integrals in a 3-index tensor format (via DF or Cholesky decomposition).
    \item \textbf{CISD trial wavefunctions:} For higher accuracy, a CISD wavefunction can be used as the trial state. This can be obtained either from a CI calculation or by projecting a CCSD wavefunction into the singles–doubles space, both of which give similar results. AFQMC with CISD trials is exact for all two-electron systems. Unlike CCSD, AFQMC energies are not strictly size-extensive, but deviations from exact size-extensivity are generally modest. AFQMC with a CISD trial is a highly accurate approach for treating not only organic molecules in their ground states but also transition states and systems containing transition metals (see Fig.~\ref{fig:afqmc_t_a}).
    \item \textbf{Local natural orbital (LNO) AFQMC}: We have also interfaced AFQMC with the LNO framework to enable linear-scaling calculations.\cite{kurian2023toward} The theoretical framework closely parallels that of LNO-CCSD,\cite{rolik2011general} and the convergence of energies with respect to LNO thresholds is similar to CCSD and CCSD(T). At present, an LNO version of AFQMC with a CISD trial is not available, since the lack of strict size-extensivity makes it nontrivial to construct a local formulation. Work to address this issue is currently in progress.
\end{enumerate}

\subsection{Excited state theories} 
Electronic excitation methods are central to quantum chemistry and materials science, enabling predictions of spectra, photochemical reactivity, and electronic transport.\cite{dreuw:2005:chemrev,onida:2002:rmp} 
Accurate treatment of excited states is essential for chromophores, catalytic intermediates, defects in solids, and optoelectronic materials. 
Methodologically, excitations are challenging: they require a balanced treatment of electron correlation effects in neutral or charged open-shell states, and need efficient scaling for large or periodic systems. 
Since version 1.9, \pyscf has substantially expanded its capabilities for simulating electronically excited states, enabling the treatment of larger chemical systems and a wider array of excited-state properties.

\begin{figure}[h!tbp] 
\begin{center}
  \subfloat[Geometry]{%
      \includegraphics[width=0.21\textwidth]{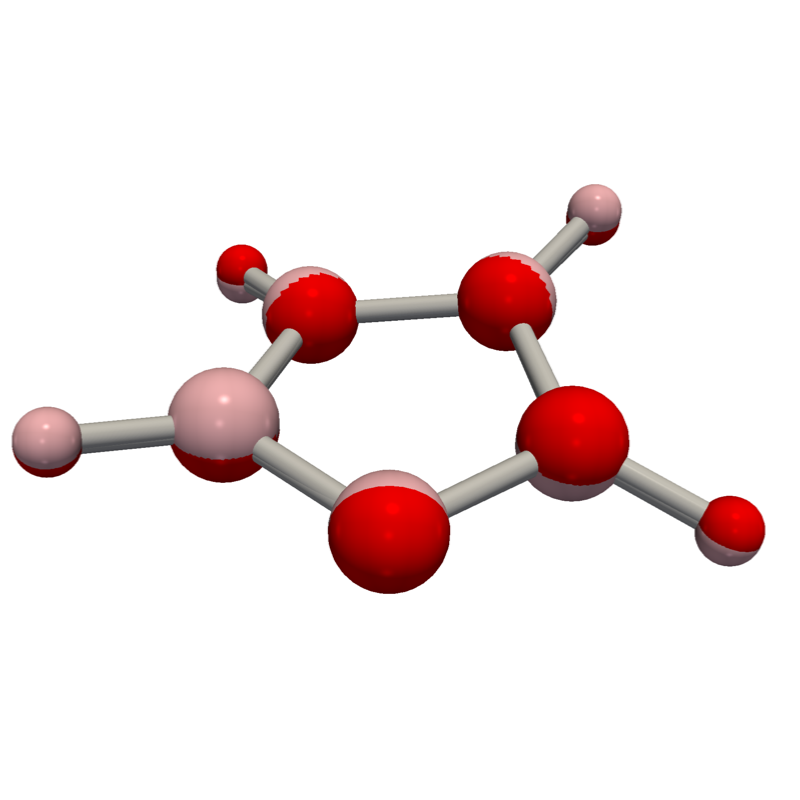}%
  }
  \hfill
  \subfloat[TDA PES in the branching plane]{%
      \includegraphics[width=0.39\textwidth]{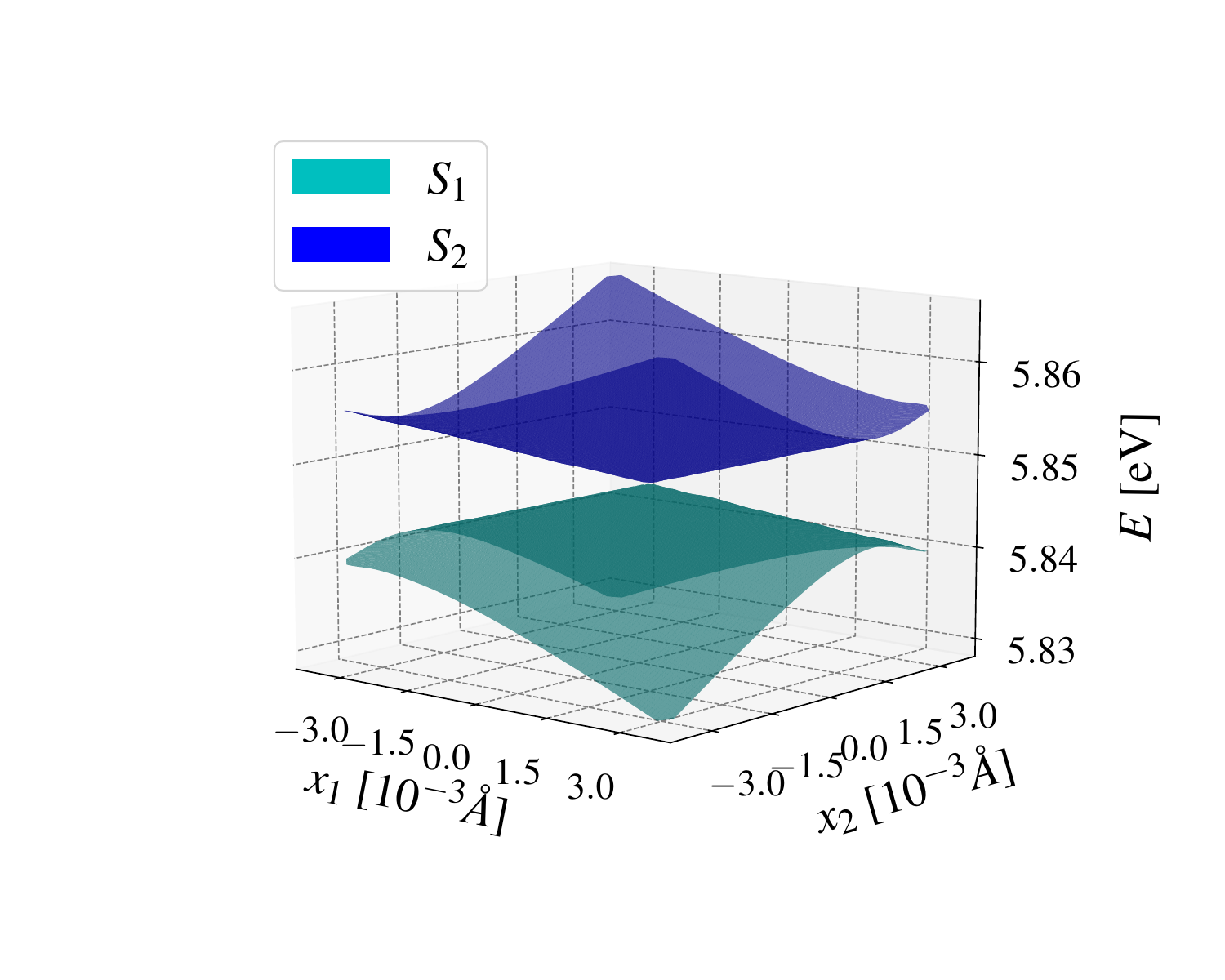}%
  }
  \hfill 
  \subfloat[TDA-ris PES in the branching plane]{%
      \includegraphics[width=0.39\textwidth]{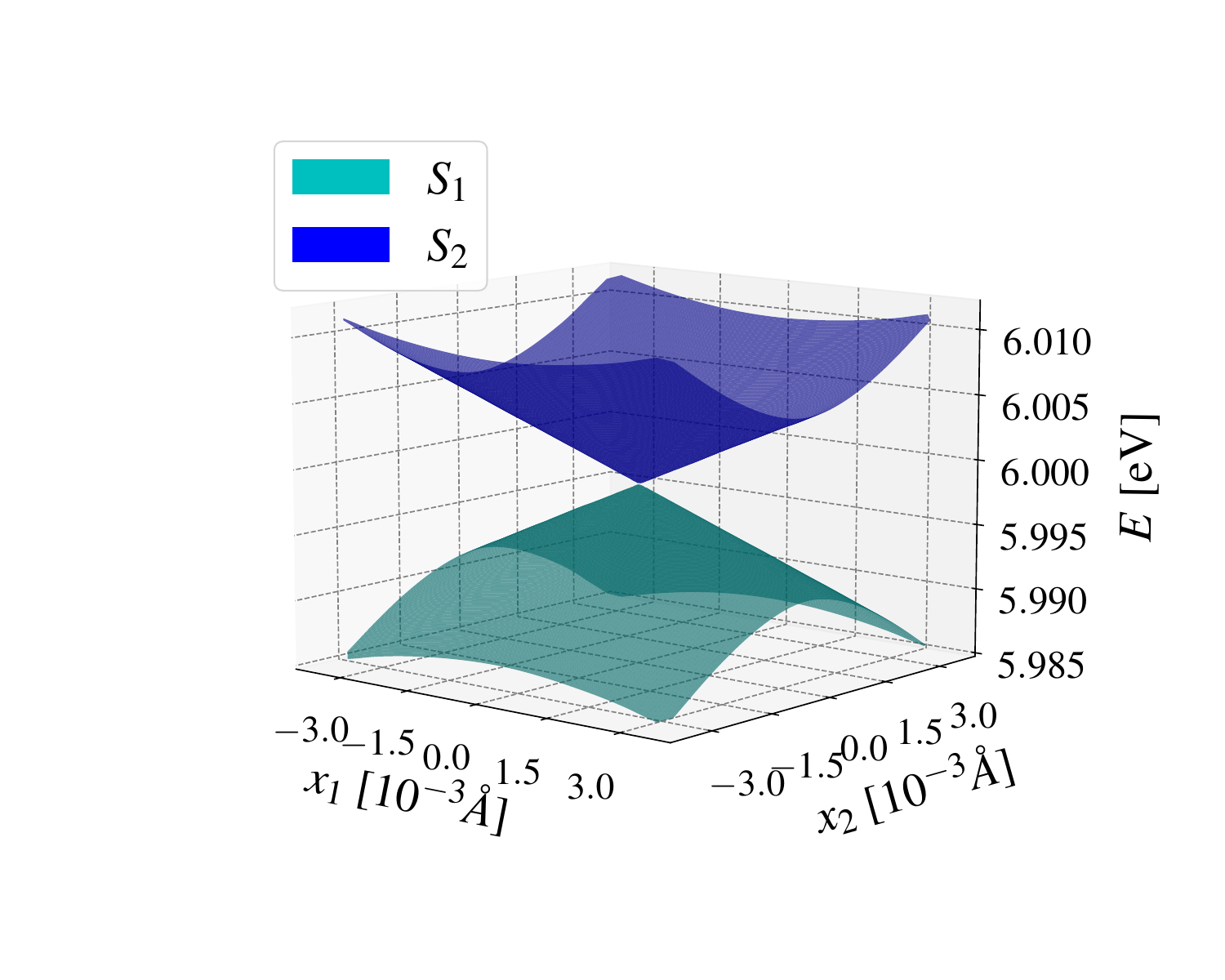}%
  }
\end{center}
\caption{
Capabilities of the TDDFT derivative couplings in \pyscf.
(a) The geometry of the $S_1/S_2$ minimum energy crossing point of furan using TDA (pink atoms) and TDA-ris (red atoms).
(b) and (c) Potential energy surfaces within the branching plane of the $S_1/S_2$ minimum energy determined using the TDA and the TDA-ris methods, respectively.
Figures are adapted with permission from Ref.~\citenum{pu2026}. Copyright \copyright 2026, American Chemical Society.
}
\label{fig:furan_mecp}
\end{figure}

Among the most widely used and computationally affordable approaches, TDDFT and its variants have become considerably more robust within \pyscf.\cite{rungegross:1984:prl} 
The package now supports linear-response and spin-flip TDDFT as well as Tamm-Dancoff (TDA) and random-phase (RPA) approximations with a wide range of density functionals for both restricted and unrestricted references.\cite{marques:2004:arpc,casida:1995p155,baurenschmitt:1996:cpl} 
These implementations enable efficient calculations of excitation energies and give access to transition dipole moments, oscillator strengths, and excited-state densities. 
Analytic excited-state gradients are also available, enabling geometry optimizations and dynamics on excited-state potential energy surfaces.\cite{furche:2002:jcp} 
Major developments have been made in the TDDFT code for periodic systems, including efficient treatments of excitons with finite momentum transfer, which are critical for spectra of solids.\cite{casida2009time}
Integration of TDDFT with continuum solvation models (C-PCM\cite{barone1998quantum} and IEF-PCM\cite{cances1997new,mennucci1997evaluation}) extends its applicability to molecular systems in realistic solvent environments.\cite{tomasi:2005:chemrev}
\pyscf also now contains a TDDFT-ris\cite{zhou2023minimal} implementation, which uses a minimal auxiliary basis set to approximate the Coulomb and exchange-type integrals, achieving substantial speedups compared to the conventional TDDFT implementation. 
To support non-adiabatic molecular dynamics, derivative couplings between the ground and excited states, as well as among excited states, have been implemented for both 
TDDFT and TDDFT-ris methods (see Fig.~\ref{fig:furan_mecp}).\cite{pu2026} 

\begin{figure}[t!]
\begin{center}
	\includegraphics{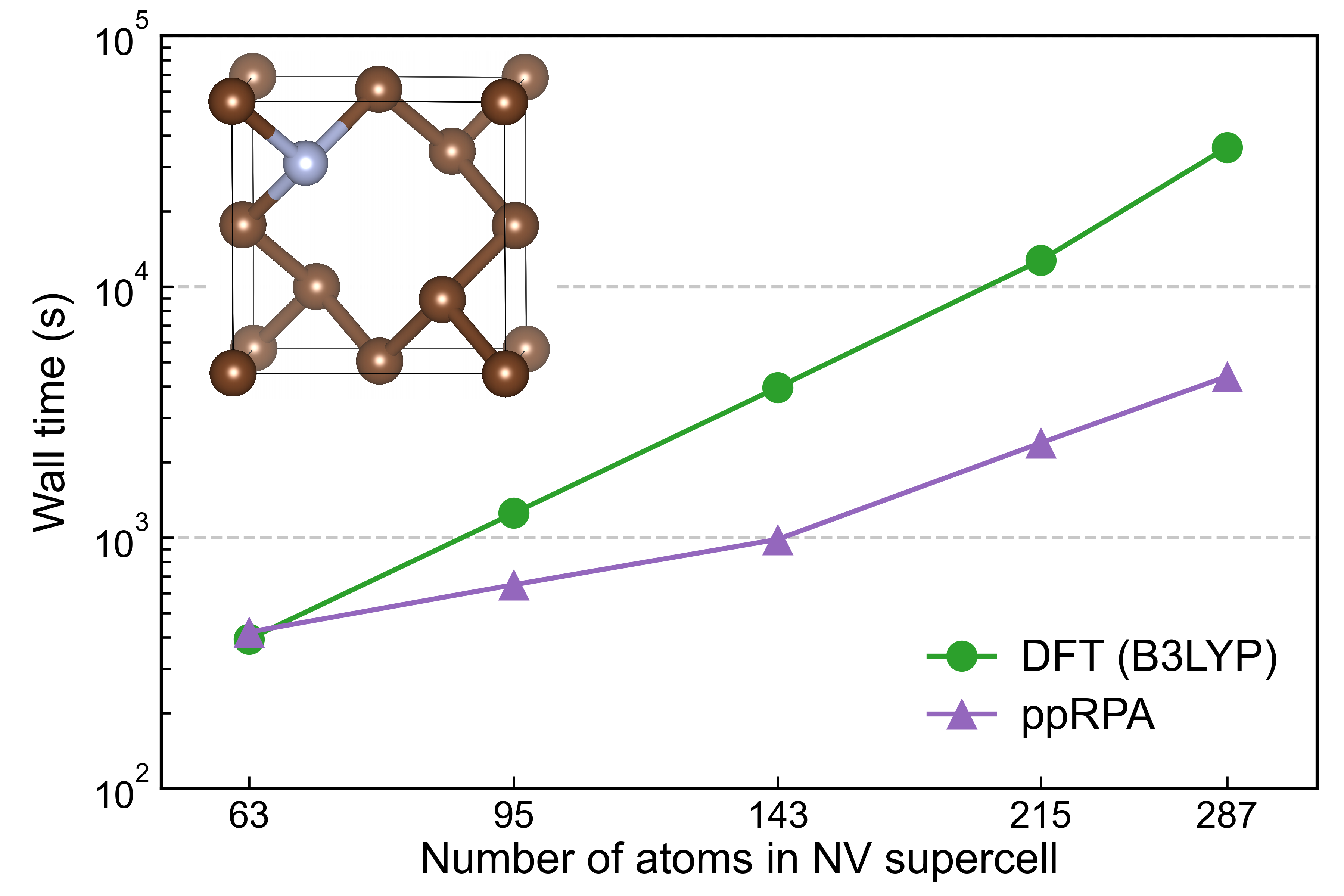}
\end{center}
\caption{Wall times for the NV center DFT (B3LYP\cite{b3lyp,libxc}) ground-state  and ppRPA excited-state calculations under $\Gamma$-point periodic boundary conditions on a 48-core CPU node. All ppRPA calculations employed an active space of 400 orbitals and Gaussian density fitting.
}
\label{fig:pprpa}
\end{figure}

\pyscf now also supports the particle–particle random phase approximation (ppRPA), extending its excited-state and correlation capabilities beyond the traditional particle–hole RPA framework.\cite{yangDoubleRydbergCharge2013,aggelen:2014p18A511} 
The ppRPA module provides access to neutral excited states in both spin-restricted and unrestricted molecular and periodic systems (without $k$-point sampling), using a density-fitting implementation with optional active-space truncation scheme built upon DFT ground-state solutions,\cite{li:2023p7811} enabling efficient calculations of valence, double, charge-transfer, Rydberg, and spin-defect excitations (see Fig.~\ref{fig:pprpa}).\cite{liAccurateExcitationEnergies2024,liParticleParticleRandom2024,yuAccurateEfficientPrediction2025}
Key property capabilities, including oscillator strengths, natural transition orbitals, analytic gradients, and relativistic corrections, are also available.

\begin{figure}[t!]
\begin{center}
	\includegraphics[width=0.95\textwidth]{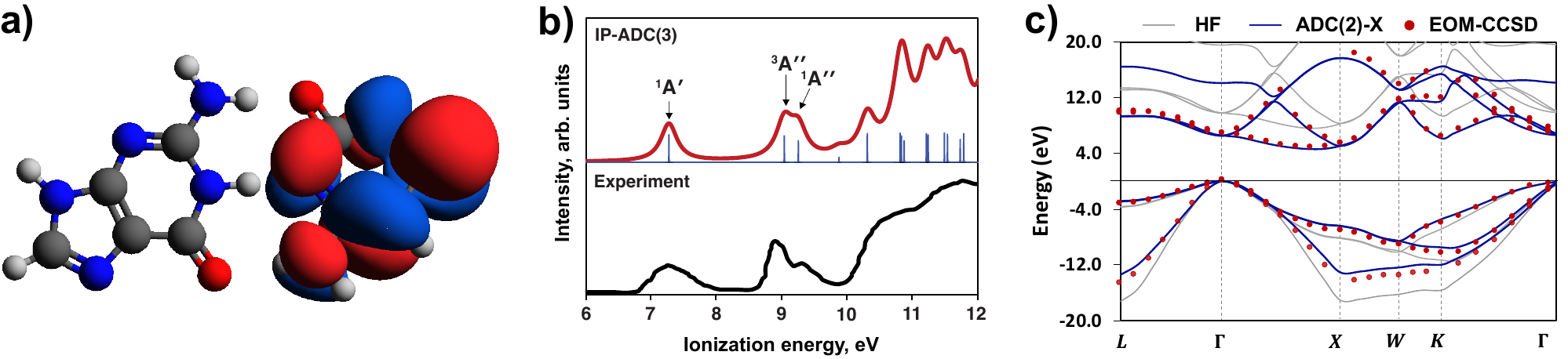}
\end{center}
\caption{
Capabilities of the ADC module in \pyscf. 
(a) Dyson orbitals for the electron-attached guanine-cytosine base pair computed using ADC(3).\cite{banerjee:2021p074105}
(b) Photoelectron spectrum of the TEMPO radical computed using ADC(3)\cite{banerjee:2021p074105} and compared to the experimental results.\cite{kubala:2013p2033}
(c) Band structure of diamond crystal computed using the periodic ADC(2)-X method\cite{banerjee:2022p5337} in comparison to HF and EOM-CCSD.\cite{mcclain:2017p1209}
Panels (a) and (b) adapted with permission from Ref.~\citenum{banerjee:2021p074105}. Copyright (2021) AIP Publishing.
Panel (c) adapted with permission from Ref.~\citenum{banerjee:2022p5337}.
Copyright (2022) American Chemical Society.}
\label{fig:adc}
\end{figure}

Beyond density-functional approaches, \pyscf also includes a comprehensive framework for algebraic diagrammatic construction (ADC) theory.\cite{schirmer:1982p2395,dreuw:2014p82,banerjee:2023p3037}
The ADC module supports calculating neutral excitation energies, ionization potentials, and electron affinities, providing a unified description of a broad range of spectroscopic processes. 
In addition to excitation energies, the ADC implementation delivers Dyson orbitals (Fig.~\ref{fig:adc}a), spectroscopic amplitudes, and one-particle density matrices, enabling detailed characterization of excited states and their properties.\cite{banerjee:2023p3037} 
Open-shell references are fully supported, broadening the applicability to radicals and other species with unpaired electrons (Fig.~\ref{fig:adc}b).\cite{banerjee:2019p224112,banerjee:2021p074105,stahl:2022p044106,stahl:2024p204104} 
Importantly, ADC has also been extended to periodic systems for charged excitations, making it possible to compute band structures and photoelectron spectra of crystalline materials (Fig.~\ref{fig:adc}c).\cite{banerjee:2022p5337,ahmed:2025p7588} 

Furthermore, \pyscf provides a comprehensive suite of $GW$ implementations that cover a wide range of numerical and self-consistency strategies for calculating quasiparticle energies.\cite{hedin:1965:pr,aryasetiawan:1998:rpp} 
In addition to the exact $GW$ formalism obtained by solving for the direct RPA excitations, \pyscf supports full-frequency numerical integration schemes based on analytic continuation and contour deformation for both molecular and periodic systems with $k$-point sampling.\cite{zhuAllElectronGaussianBasedG0W02021} 
\pyscf also features recently developed frequency-free $GW$ formalisms that recast $GW$ as a supermatrix eigenvalue problem and leverage its explicit connection to equation-of-motion coupled-cluster theory.\cite{bintrim:2021p041101,tolle:2023p124123} 
These complementary implementations allow users to balance accuracy and efficiency across core and valence excitation regimes, while enabling new theoretical developments such as fully analytic $G_0W_0$ gradients.\cite{tolle:2025p3672} 
\pyscf offers multiple flavors of self-consistent $GW$ implementations for molecular and periodic systems, from quasiparticle self-consistent $GW$ (QSGW) and eigenvalue self-consistent $GW$ (evGW) to fully self-consistent $GW$ (SCGW, molecular only).\cite{vanschilfgaarde:2006:qsgw,leiGaussianbasedQuasiparticleSelfconsistent2022} 
Beyond quasiparticle calculations, \pyscf now supports Bethe-Salpeter equation (BSE) calculations of neutral excitation energies and optical spectra for molecular systems, in combination with the aforementioned $GW$ formalisms.\cite{rohlfing:2000:bse} 
An energy-specific Davidson algorithm is implemented to enable efficient $GW$+BSE simulations of core excitations and dense excited-state manifolds.\cite{hillenbrand:2025:es-bse} 
Beyond ADC and $GW$, \pyscf also offers a variety of many-body Green's function implementations for calculating one-particle spectral functions at the full CI, EOM-CCSD, and second-order Green's function theory (GF2) levels.\cite{stanton:1993:jcp,phillips:2014:gf2,Laughon2022ccgf}

At the high-accuracy end, \pyscf offers equation-of-motion coupled-cluster methods with single and double excitations (EOM-CCSD).\cite{stanton:1993:jcp,bartlett:2012p126} 
Several flavors of methods are available, including EOM-CCSD for neutral excitation energies, spin-flip, ionization potentials, and electron attachments.\cite{krylov:2008:arpc}
EOM-CCSD methods have also been generalized to periodic boundary conditions, enabling the treatment of quasiparticle and excitonic excitations in solids at the CC level.\cite{mcclain:2017p1209,wang:2020p3095,wang:2021p6387,vo:2024p044106} 
Improvements in solver stability, memory usage, and parallel scalability further enhance the reliability of EOM-CC across both molecular and extended systems. 

Alongside these methodological innovations, the infrastructure has also advanced in ways that directly benefit excited-state calculations. 
More efficient eigensolvers and DF techniques have reduced computational cost and improved robustness.\cite{weigend:2002:ri} 
Transition properties are now handled consistently across different methods, simplifying workflows that combine multiple levels of theory. 
Emerging GPU acceleration and improved parallelization are also beginning to reach the excited-state modules, laying the foundation for substantial performance gains on modern architectures.\cite{li:2025:gpu4pyscf}

\subsection{Multireference wavefunction methods and multireference density functional theory}
A multireference electronic structure model is required for qualitatively accurate simulations of electronic states characterized by strong static correlation. Since the early versions, \pyscf has had extensive support for multireference wavefunction methods, such as MCSCF, NEVPT2, and (via interfaces to {Dice}\cite{smith2017cheap} and \textsc{block2}\cite{zhai2023block2}) selected configuration interaction, DMRG, MRCI, and other methods. 

The state-averaged multi-configurational
self-consistent field (SA-MCSCF) implementation now supports analytical nuclear
gradients and non-adiabatic couplings (NACs) between states. In addition, driven
similarity renormalization group (DSRG)\cite{evangelista2014driven} methods and the second-order
multireference perturbation theory (DSRG-MRPT2)\cite{li2015multireference} have been incorporated.
\pyscf offers automatic active-space construction schemes for complete active space
self-consistent field (CASSCF) calculations. In addition to the atomic valence active space (AVAS)\cite{sayfutyarova2017automated}
method, the approximate pair coefficient (APC)\cite{King2021a} approach has been added to the
package for the automation of active-space selection in high-throughput
workflows. Several multi-reference correlation methods based on the
non-orthogonal configuration interaction (NOCI) formalism are implemented, such
as spin-flip non-orthogonal configuration interaction with the grouped-bath ansatz (SF-GNOCI)\cite{park2025efficient} and multi-state
DFT (MSDFT).\cite{bao2020block}

At the same time, all many-electron systems of any type exhibit quantitative dynamical correlation effects. The current version of \pyscf brings new capabilities to include such effects via the multiconfiguration pair-density functional theory (MC-PDFT)\cite{Manni2014} approach, enabling the modeling of highly multiconfigurational electronic states with accuracy comparable to second-order multireference perturbation theory at a much lower computational cost.

\pyscf implements MC-PDFT according to the protocol proposed by Li Manni {et al.},\cite{Manni2014} in which the density and pair density are evaluated for a multiconfigurational wave function taken without modification from an underlying MCSCF calculation. The implementation is compatible with any number of electronic states and any state-averaging scheme available in \pyscf. It is also compatible with modified full CI solvers or approximations to the active-space full CI subproblem such as DMRG, provided the corresponding 1- and 2-body reduced density matrices are available (with the caveat that modifications of this type will usually disable analytical energy gradient support; see below). Additionally, because na\"{i}ve MC-PDFT models of nearly-degenerate electronic states are often qualitatively inaccurate in the immediate vicinity of conical intersections or locally avoided crossings, \pyscf implements a variety of ``multi-state'' MC-PDFT methods, of which the most significant is linearized pair-density functional theory (L-PDFT).\cite{Hennefarth2023}

The ``on-top'' density functional part of the MC-PDFT energy expression is evaluated using either ``translated''\cite{Manni2014} or ``fully-translated''\cite{Carlson2015} density functionals specified (in most cases) by appending a ``t'' or ``ft'' prefix, respectively, to the name of an existing KS-DFT exchange-correlation functional available in Libxc.\cite{libxc} Any (pure or global hybrid) LDA or GGA functional is compatible with either translation scheme, and (pure or global hybrid) meta-GGA functionals which uses the kinetic energy density (rather than the Laplacian) are compatible with the original translation scheme.\cite{Bao2025}

Analytical MC-PDFT energy gradients are implemented using the technique of Lagrange multipliers. The Lagrangian depends directly on the error function whose minimization determines the underlying wave function, so any variant of the underlying wave function requires a unique implementation. In \pyscf, analytical gradients for MC-PDFT (single or multi-state) energies are available for calculations based on underlying CASSCF or state-averaged CASSCF calculations. In addition, analytical gradients and NAC vectors are available for some specific types of multi-state PDFT calculations using some functional types, which are summarized in Table \ref{tab:mcpdft_grad_capabilities}. In addition to this, spin–orbit effects for MC-PDFT and L-PDFT\cite{jangid2025linearized} can be included via a two-step state-interaction approach, and additional molecular properties (e.g., analytical dipole moments\cite{clifford2025analytic}) have also been implemented.

\begin{table}
\caption{\label{tab:mcpdft_grad_capabilities}The highest available rung of functionals for which a given feature of MC-PDFT is available for wave functions of various types as of \pyscf v2.12. The possibilities are None, LDA, GGA, and mGGA. XMS-PDFT, CMS-PDFT, and L-PDFT are various types of multi-state PDFT methods.}
    \begin{tabular}{c|ccc}
    \hline\hline
    Wave function type & Energy & Gradient & NAC \\\hline
    State specific & mGGA & mGGA & None \\
    State-averaged & mGGA & mGGA & None \\
    XMS-PDFT\cite{Bao2020a} & mGGA & None & None \\
    CMS-PDFT\cite{Bao2020,Bao2022,Calio2024} & mGGA & mGGA & mGGA \\
    L-PDFT\cite{Hennefarth2023} & mGGA & GGA & None \\
    \hline\hline
    \end{tabular}
\end{table}

\subsection{Molecular dynamics}

While the energy and force outputs of \pyscf can easily be fed into many external molecular dynamics packages,\cite{seal_computing_2025} 
the molecular dynamics (MD) module in \pyscf enables \textit{ab initio} Born--Oppenheimer molecular dynamics simulations to be fully integrated into the \pyscf's quantum chemistry framework. It supports the $NVE$ ensemble using a velocity Verlet integrator and the $NVT$ ensemble using a Berendsen thermostat (more sophisticated thermostats are being implemented). 
For thermal initial conditions, the module can sample atomic velocities from a Maxwell--Boltzmann distribution at a user-specified temperature.

This MD capability can be used with any electronic structure method in \pyscf that provides analytic gradients, to propagate nuclei on their corresponding potential energy surfaces. This allows one to compute spectroscopic and thermodynamic properties and probe reaction dynamics with the plethora of electronic structure methods in \pyscf. The NVE module provides a practical route to check stability of active spaces for dynamics with active-space based multi-reference electronic structure such as CASSCF and MC-PDFT.\cite{seal_weighted_2025} Besides traditional \textit{ab initio} molecular dynamics, the MD module serves as a platform for testing machine-learning-based approaches that leverage intermediate electronic-structure information for dynamical simulations\cite{rath_interpolating_2025} as well as for generating training datasets for machine-learned potentials. Overall, by integrating MD functionality into the \pyscf ecosystem, the module enables \textit{ab initio} molecular dynamics and the development of hybrid quantum-machine learning approaches within a single Python-based platform.

\subsection{Solvation models and QM/MM}
Implicit solvent models provide a practical way to include solvation effects in quantum chemistry. They are essential when one wants to obtain chemically realistic results in solution-phase chemistry without the prohibitive cost of full explicit solvent simulations. By mimicking the dielectric response of the solvent, implicit solvent models give realistic energies and geometries close to experiment. These models also provide estimates of solvation free energies and binding affinities. Since \pyscf 2.6.0, the implicit solvent module has been significantly enhanced. Besides dd-COSMO\cite{cances2013domain} and dd-PCM,\cite{stamm2016new} four PCM-type models, C-PCM,\cite{barone1998quantum} IEF-PCM,\cite{cances1997new,mennucci1997evaluation} COSMO,\cite{klamt1993cosmo} SS(V)PE,\cite{chipman1997charge,chipman2002comparison} including their analytical gradient and analytical Hessian have been added. For the detailed description of the four newly-added models, we refer the reader to Ref.~\onlinecite{herbert2021solvent}. The SMD model \cite{SMD2009} which describes non-electrostatic terms with the solvent-accessible surface area (SASA) and atomic surface tension is also supported.

Solving the implicit solvent model is usually cheaper than a regular SCF iteration. In the \pyscf implementation, a direct solver is used for the linear system associated with the solvent models. Therefore, the time and space complexity of the solvent models are $O(N^3_g)$ and $O(N^2_g)$ respectively where $N_g$ is the number of grid points on the molecular surface. IEF-PCM and SS(V)PE need several times more memory than C-PCM and COSMO. These implementations are designed for simulating the solvation effect of smaller molecules ($<$ 100 atoms). For larger molecules, dd-COSMO and dd-PCM methods are recommended. Unfortunately, the analytical Hessians of the more scalable solvers dd-COSMO and dd-PCM methods are still unavailable. The solvent models can be used with restricted/unrestricted HF and DFT with or without density fitting. TDDFT and post-HF methods with solvents\cite{Cammi1999,tomasi:2005:chemrev,herbert2021solvent} are also supported as mentioned earlier. These modules are also GPU-accelerated in \gpupyscf (see below) except for the non-electrostatic terms in the SMD model which are computationally inexpensive, for which \pyscf's implementation is based on the open-source code from NWChem.\cite{nwchem}

\begin{figure}
    \centering
    \includegraphics[width=0.65\linewidth]{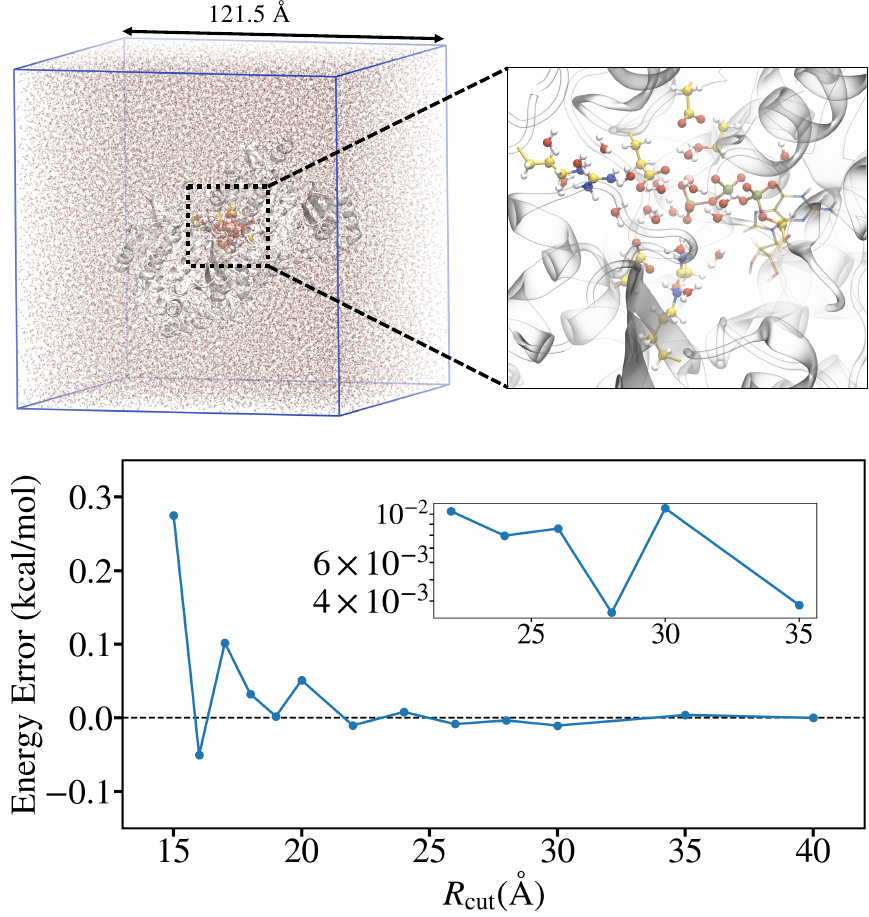}
    \caption{QM/MM energy convergence in a model for GTP hydrolysis in microtubules. The upper panel shows the model setup with the QM region shown as spheres and the MM region (protein and water) shown as sticks or a cartoon in the upper left. A zoom-in QM region is shown in the upper right. The bottom panel shows the QM/MM energy convergence with an increasing cutoff distance for the short-range multipolar components, i.e., multipoles up to quadrupoles, with an inset zoom-in. Copyright \copyright 2025, American Chemical Society.}
    \label{fig:qmmm}
\end{figure}

In addition to implicit solvent models, \pyscf now supports periodic QM/MM simulations of complex, heterogeneous condensed-phase chemistry. This implementation embeds a QM subsystem in a periodic box of classical (MM) point charges or Gaussian-distributed charges. The long-range electrostatic interactions---both between the QM region and its periodic images and between the QM region and the infinite MM lattice---are treated rigorously using an efficient multipolar Ewald approach with controllable errors (see Fig.~\ref{fig:qmmm}).\cite{li2025accurate} Analytic gradients are available for both QM and MM atomic positions, enabling geometry optimizations and, in conjunction with external molecular dynamics packages,\cite{LAMMPS,ipi} simulations of both classical and quantum nuclear dynamics on the QM/MM potential energy surface. Furthermore, GPU acceleration of all the QM/MM-related computational bottlenecks is supported through fast electron integrals by \gpupyscf and tensor contractions by {CuPy}\cite{cupy_learningsys2017} or {cuTENSOR}.

\subsection{GPU4PySCF}
\label{sec:gpu4pyscf}
GPU devices are now widely available through supercomputers and cloud computing environments.
To make effective use of these resources, the \gpupyscf extension was developed to accelerate \pyscf computations.
The focus of the \gpupyscf extension differs from that of \pyscf.
In \pyscf, the emphasis is on conservative parameters and algorithms to ensure broad functionality, while \gpupyscf prioritizes optimized performance.

\gpupyscf has fewer features than \pyscf.
Its main functionalities focus on DFT and DFT-related methods.
For post-HF methods, \gpupyscf (as of version 1.4.2) only supports
the integral-in-core algorithms for restricted MP2 and CCSD.
Most of the features in \gpupyscf run on a single GPU.
Some modules can use multiple GPUs on the same node to access more GPU memory.
However, multi-GPU parallelization has not been fully optimized, and the scaling efficiency is relatively low.
\gpupyscf can be installed from the Python Package Index (PyPI) using precompiled wheels, which currently support only NVIDIA GPUs.
Table~\ref{tab:gpu4pyscf:feat} lists the main features in \gpupyscf and their approximate speedups compared to the CPU performance of \pyscf.

\begin{table}
  \centering
  \caption{Main features of \gpupyscf. The performance speedups relative to \pyscf
  are estimated on a single NVIDIA A100 GPU.}
  \label{tab:gpu4pyscf:feat}
\begin{tabular}{lllllllll}
\hline\hline
  Features                                        & speedup vs \pyscf \\
\hline
  Density fitting DFT (energy, gradient, Hessian) & $> 1,000$   \\
  Integral-direct DFT (energy, gradient, Hessian) & 500 -- 1,000 \\
  TDDFT (energy, gradient, non-adiabatic coupling)                        & $\sim$ 500  \\
  PCM solvent model (energy, gradient, Hessian)   & $> 1,000$   \\
  Infrared and Raman spectroscopy                 & 500 -- 1000  \\
  Polarizability                                  & $\sim$ 500  \\
  NMR shielding                                   & $\sim$ 100  \\
  Energy decomposition analysis                   & ---$^a$       \\
  DFT with Periodic boundary condition            & 300 -- 500   \\
  PBC-DFT gradients and stress tensor             & 200 -- 300   \\
\hline\hline
\multicolumn{2}{l}{$^a$ Not available in \pyscf}
\end{tabular}
\end{table}

\gpupyscf can achieve speedups of 2--3 orders of magnitude over the \pyscf CPU-based code.\cite{wu2025}
The most notable performance advantage in \gpupyscf is found in the density fitting (DF) module.
The DF integral approximation involves many tensor contractions and is highly floating-point operation (FLOP)-intensive.
Due to the substantial advantages of tensor contraction operations on GPUs,
DF in \gpupyscf delivers substantial speedups for DFT tasks, including the
computation of ground state energies, nuclear gradients, nuclear Hessians, and
time-dependent DFT (TDDFT).
For medium-sized molecules (50--100 atoms), DF-DFT calculations in \gpupyscf
show the largest performance gains, and in some cases
a single NVIDIA A100 GPU can outperform 1,000 CPU cores.

The integral algorithms for Gaussian-type orbitals in \gpupyscf have been intensively optimized for GPU hardware.
Although the integral-direct approach cannot fully utilize the GPU compute units as well as the
density-fitting methods, integral-direct DFT still achieves nearly 1000 times speedup compared to the \pyscf CPU implementation.
For large molecules, such as those with more than 5,000 basis functions, the
integral-direct approach performs particularly well.
\gpupyscf can routinely handle integral-direct DFT calculations with 30,000 basis functions on a single 80 GB VRAM GPU.

To make the user experience of \gpupyscf similar to that of \pyscf, the APIs in \gpupyscf generally follow those of \pyscf.
\gpupyscf uses the same naming conventions, module structure, and function signatures as in \pyscf.
It also provides Python classes for individual quantum chemistry methods.
In many cases, the \gpupyscf features can be imported in the same way as in \pyscf,
with the only difference being that \verb$pyscf$ is replaced with \verb$gpu4pyscf$ in the import statements.


To make it easier to use \gpupyscf features in a \pyscf program, two methods,
\verb$.to_gpu()$ and \verb$.to_cpu()$, are provided by compatible \pyscf and \gpupyscf classes, respectively.
The \verb$.to_gpu()$ method in \pyscf classes converts a \pyscf instance and all its attributes to the corresponding \gpupyscf objects.
For more details on how to use \gpupyscf, we refer to Ref.~\onlinecite{pu2025}, which is a recently written tutorial.

\subsection{\pyscfad}

\pyscfad\cite{pyscfad} is a separate companion module to {\pyscf} which provides a fully differentiable in-core reimplementation of \pyscf, and leverages modern advances in automatic differentiation (AD) to make property calculations and response theory more accessible.
The design philosophy mirrors that of the core \pyscf,
which is to enable rapid development of new ideas while allowing benchmarks on properties
that were difficult to compute due to the lack of analytic derivatives.
The project is openly developed on GitHub,\cite{pyscfadrepo} and the package is distributed via PyPI.

Historically in quantum chemistry method development, 
analytic derivatives have often lagged far behind the corresponding implementation of energy evaluations. For example, analytic derivative couplings\cite{send2010first,li2014first,ou2015first,zhang2015analytic} in linear response TDDFT with Gaussian type orbital basis were developed over two decades after the original energy formalism.\cite{tddft,casida1995time} 
Although mathematically trivial, manually deriving analytic derivatives is tedious and error-prone, especially for higher-order derivatives and when the wavefunction response is involved. Fortunately, modern AD tools remove much of this difficulty, yet most quantum chemistry codes remain disconnected from such infrastructures. \pyscfad~serves to bridge this gap.

\pyscfad~supports both \numpy~and \jax\cite{jax,jaxrepo} backends, with the latter providing differentiability and hardware acceleration via JIT compilation.
A subset of methods in \pyscf~are included, namely,
HF, DFT, MP2, RPA, CC, orbital localization methods, etc.
Most of them allow up to fourth-order nuclear derivatives and first-order derivatives with respect to basis function parameters. Higher-order derivatives can
be achieved by a simple extension to the underlying electron integral engine {Libcint}.\cite{libcint} Example applications of \pyscfad~in a variety of quantum chemistry computational tasks have been presented,\cite{pyscfad,lnoccad} highlighting its flexibility and extensive functionality.

\begin{figure}[t!]
\begin{center}
	\includegraphics[width=.95\textwidth]{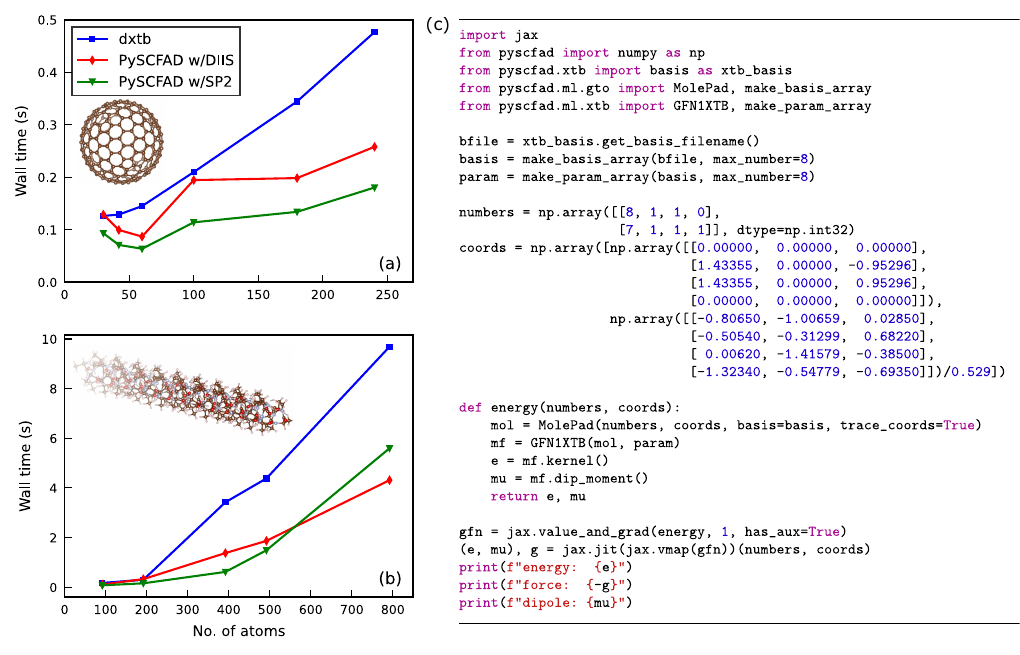}
\end{center}
\caption{Wall times of GFN1-xTB energy and nuclear gradient calculations for fullerenes (a) and alanine polypeptides (b) using \textit{dxtb} (blue) and \pyscfad with the DIIS\cite{pulay1980convergence} (red) and SP2\cite{niklasson2002expansion} (green) solvers. 
The molecular geometries were taken from Ref.~\citenum{steinbach2025acceleration}, and the calculations were performed on a single NVIDIA A100 GPU.
(c) An example of a batched xTB calculation using \pyscfad.}
\label{fig:xtb}
\end{figure}

\pyscfad~preserves a similar API to \pyscf~while introducing the \verb|pyscfad.numpy| module,
which provides a unified interface for functions across different \numpy-like backends.
In addition, it offers a generic framework for implicit differentiation
of iterative solvers,\cite{pyscfad}
which is traditionally accomplished by solving 
the coupled-perturbed Hartree-Fock equations\cite{pople1979derivative}
or employing the $Z$-vector approach.\cite{handy1984evaluation}
Adding new methods to \pyscfad is straightforward, with the primary requirements that functions be pure to ensure compatibility with JIT compilation and that array shapes be static for optimal performance.
Recently, the GFN1-xTB\cite{grimme2017robust} method was implemented within \pyscfad, achieving performance that surpasses \textit{dxtb}\cite{friede2024dxtb} in most cases [see Fig.~\ref{fig:xtb}(a) and (b)]. In addition, our implementation naturally benefits from \jax's automatic vectorization, enabling efficient batched calculations [see Fig.~\ref{fig:xtb}(c)].

As \pyscfad~remains under active development, several future directions are envisioned:
(1) Incorporating efficient electron integral kernels from \textsc{GPU4PySCF} to achieve better performance. Currently, integral evaluation is restricted to CPUs, although the rest of the program already benefits from \jax's native hardware acceleration.
(2) Improving interoperability to enable drop-in use within computational workflows beyond quantum chemistry. As a preliminary demonstration, we have integrated the prediction of an effective Hamiltonian with \pyscfad~to explore the design space of machine-learning models.\cite{suman2025exploring}

\subsection{Symmetry}


The treatments of point-group symmetry and spin symmetry have been  improved in several modules. By integrating the \texttt{libmsym} library,\cite{johansson2017automatic} \pyscf now supports the
generation of symmetry-adapted real orbitals in terms of the spherical
atomic orbitals for any point group.
In the full configuration interaction (FCI) solvers,
cylindrical symmetry can be imposed on the FCI wavefunctions, while
a configuration state function (CSF) formulation has also been developed to enable
spin-symmetry-adapted FCI calculations.
In addition, space-group symmetry is natively supported in \pyscf with PBCs 
(an optional interface to \texttt{spglib}\cite{spglib} is also provided), 
which enables space-group symmetry-adapted calculations 
for mean-field, MP2, and CC methods with $k$-point sampling.

\subsection{Interoperability}

\pyscf provides multiple interfaces and standardized data formats to facilitate interoperability with other quantum chemistry software and simulation frameworks.
\pyscf integrates with the Atomic Simulation Environment (ASE),\cite{ase} providing energy and force calculators for molecular systems at the mean-field and post–HF levels, and stress tensors for periodic materials at the mean-field level.
For standardized data exchange, \pyscf supports the TREXIO\cite{trexio_2023} file format, allowing users to export data such as atomic positions, lattice vectors, basis sets, effective core potentials, atomic and molecular orbitals, one- and two-electron integrals, and one- and two-particle density matrices. 
\pyscf also supports inputs following the QCSchema\cite{qcschema} specification.
\pyscf further supports several file formats widely used in the quantum chemistry community, including Molden files, VASP CHGCAR files, Gaussian cube files, and FCIDUMP files.
In addition, basis sets now can be parsed directly from the Basis Set Exchange\cite{bse} repository.
Finally, \pyscf implements export utilities for downstream COSMO-RS\cite{klamt1998refinement,gerlach2022open} workflows, generating COSMO files in the Turbomole\cite{turbomole} format from conductor-limit PCM calculations.

\section{Conclusions}

After ten years of development, \pyscf is a mature project that forms part of the essential cyberinfrastructure of quantum chemistry, materials modeling, machine learning, and quantum information science. In the coming years, the developers of \pyscf aim to build on both the successes of this project as well as the lessons learnt from open-source development at scale. We envision continued expansion into new directions, such as improving the support for emerging GPU hardware and infrastructure, as well as an increasing focus on low-precision computation. In the context of new programming paradigms and machine learning, we anticipate further development of \pyscfad and utilization of JIT compilation, as well as the possibility of closer integration with emerging ML models and semi-empirical methods. New types of linear scaling methods, especially for high accuracy quantum chemistry in materials, will continue to be developed, as well as new improvements of mean-field approaches based on advances in numerical representations. We can expect additional progress in the treatment of electron correlation, solvation, excited states, and the many traditional modeling tasks of quantum chemistry. And finally, we can expect more functionality for molecular dynamics on \textit{ab initio} quality potential energy surfaces.

Above all, the last decade of development of \pyscf has relied on the project's ability to attract and support a dedicated set of community developers, who are now collectively the authors of the \pyscf package. The lifespan and activity of an open-source project is determined entirely by its community. We look forward to new ways to engage, support, and expand the \pyscf community for many years to come.

\section{Acknowledgments}

The primary writing of this article was supported by the US National Science Foundation via the NSF CSSI grants OAC-2513474 (G.K.-L.C., S.S.), OAC-2513473 (T.Z.), OAC-2513475 (L.G., M.R. Hermes), OAC-2513476 (T.C.B.), and from the US National Science Foundation under Grant No.~CHE-2044648 (A.Y.S) and the U.S.~Department of Energy, Office of Science, Basic Energy Sciences under Award No.~DE-SC0026341 (A.Y.S).
H.S.C., M.R. Hennefarth, B.J., D.K. A.O.L., T.R.S., A.S., are/were partially supported by the US DOE Office of Science, Office of Basic Energy Sciences, under Award Numbers DE-SC0023382 and DE-SC0022572, by the US National Science Foundation under Award Number CHE-2435218, and by the US AFOSR, under Award Number FA9550-20-1-0360.
The Flatiron Institute is a division of the Simons Foundation.

The author list of the current article primarily reflects major contributors since the last \pyscf paper. 
We would also like to acknowledge the following people who have contributed to earlier versions of the \pyscf codebase:
Marc Barbry,
Nick Blunt,
George H. Booth,
Allan Chain,
Matthew Chan,
Li Chen,
Jia Chen,
Yixiao Chen,
Vijay Gopal Chilkuri,
Chun-Yu Chow,
Matthias Degroote,
Jesse Gorzinski,
Sheng Guo,
Till Hanke,
Matthew R Hennefarth,
Eric Hermes,
Carlos Jimenez-Hoyos,
Konstantin Karandashev,
Muhammad Firmansyah Kasim,
Raehyun Kim,
Yusuke Kimura,
Kevin Koh,
Peter Koval,
Michal Krompiec,
Junhao Li,
Weitang Li,
Zhendong Li,
Junzi Liu,
Narbe Mardirossian,
Kazuhito Matsuda,
James D. McClain,
Mario Motta,
Bastien Mussard,
Al Nejati,
Max Nusspickel,
Artem Pulkin,
Wirawan Purwanto,
Peter Reinholdt,
Paul J. Robinson,
Elvira Sayfutyarova,
Maximilian Scheurer,
Felipe S. S. Schneider,
Henry Schurkus,
James Serna,
Pablo del Mazo Sevillano,
Unathi Koketso Skosana,
Jiace Sun,
Shining Sun,
Steffen Thomas,
Matthew Treinish,
Lucas K. Wagner,
Lee-Ping Wang,
Zhi Wang,
Oskar Weser,
Will Wheeler,
Alec White,
Sebastian Wouters,
Xin Xing,
Jun Yang,
Jason Yu,
Minye Zhang,
Boxiao Zheng, and
Zhenyu Zhu.
We also acknowledge the generous support of the general open-source \pyscf developer and user community.

%

\end{document}